# Assessing the spectral characteristics of band splitting type II radio bursts observed by CALLISTO spectrometers

Felix N. Minta [a,*], Satoshi I. Nozawa [b], Kamen Kozarev [c], Ahmed Elsaid [d,e], Ayman Mahrous [a]

[a] *Department of Space Environment, Institute of Basic and Applied Sciences, Egypt-Japan University of Science and Technology, 21934, Alexandria, Egypt.*
[b] *Department of Science, Ibaraki University, 2-1-1 Bunkyo, Mito, Ibaraki 310-8512, Japan.*
[c] *Institute of Astronomy with National Astronomical Observatory, Bulgarian Academy of Sciences, Sofia, Bulgaria.*
[d] *Department of Applied and Computational Mathematics, Institute of Basic and Applied Sciences, Egypt-Japan University of Science and Technology, Alexandria, Egypt*
[e] *Mathematics & Engineering Physics Department, Faculty of Engineering, Mansoura University, PO 35516, Mansoura, Egypt.*

**Abstract**

Metric type II radio bursts are usually early indicators of CME-driven shocks and other space weather phenomena in the solar corona. This paper presents a detailed investigation of the spectral properties of band splitting type II radio bursts and their association with sunspot number. Using type II radio bursts in a frequency range [20MHz –200MHz] observed by CALLISTO from 2010 to 2017, it was discovered that the analyzed type II shock height, magnetic field strength, CME/shock speed, and Alfvén speed synchronize with the trajectory of the solar cycle 24. Also, the study revealed that the onset of the declining phase of solar cycle 24 has the highest electron density. The analysis ascertained that the frequency of type II bursts depicts a bimodal distribution during the study period, with peaks in 2012 and 2015. Further, a good correlation (with correlation factor R = 0.87) was obtained between the estimated CME/shock speeds from the dynamic spectra and the associated CME speeds from SOHO/LASCO. Moreover, the study confirmed a significant correlation (R= ~0.8) between the absolute drift rates and the plasma frequency. Additionally, the study explored that ~60% of the type II radio bursts considered in this study emanated from the western longitudes. Hence, these findings emphasize that the temporal dynamics of the physical conditions of band splitting type II radio are essential parameters in space weather monitoring and forecasting.

*Keywords:* Radio bursts, Band splitting, Type II, CALLISTO, Solar cycle

## 1. Introduction.

Solar Radio Bursts (SRBs) are generally classified into five categories, namely [type I, type II, type III, type IV, and type V] based on their wavelengths [metric (m), deca-hectometric (DH), and kilometric (km) in the case of type II], spectra features, and associated space weather phenomena (Ramesh et al., 2013; Ratcliffe et al., 2014; Mercier et al., 2015; Carley et al., 2017; Kumari et al., 2020). Type II radio bursts produce slowly drifting, long-lasting dynamic spectra and are often associated with shocks advancing with super- Alfvénic speeds ahead of a coronal mass ejection (CME) in the solar corona. They are noted to be plasma emissions that are closely related to solar eruptive events in the solar corona (Cliver et al., 2005; Temmer et al., 2010; Grechnev et al., 2011; Vasanth et al., 2011; Schmidt et al., 2014; Kumari et al., 2017a). The availability of the Solar Heliospheric Observatory (SOHO) at the beginning of the solar cycle 23 has made it possible to ascertain the relationship between type II radio bursts and CMEs (Gopalswamy et al., 1998; Kumari et al., 2017b; Morosan et al., 2020; Majumdar et al., 2021).

* Corresponding author.
*E-mail address*: felix.minta@ejust.edu.eg (F. N. Minta)

Earlier studies pointed out that band splitting in type II radio bursts results from simultaneous radio emissions downstream and upstream of the shock (Tidman, 1965; Mann et al., 1995; Hariharan et al., 2014). According to Rankine–Hugoniot equation, the difference in frequencies between the two split bands (density jump), which correlates with the shock compression ratio, can be expressed to estimate the Alfvén Mach number (Smerd et al., 1975; Vršnak et al., 2001b). The CME heliospheric height at the type II onset time indicates the height at which the CME becomes super- Alfvénic to trigger a fast mode magneto-hydrodynamic (MHD) shock (Gopalswamy et al., 2013). Vasanth et al., (2014) also confirmed that type II radio bursts with band splitting are crucial in estimating the coronal magnetic field strength. The magnetic field significantly influences coronal heating, particle acceleration, and the generation of solar eruptive structures such as prominences and CMEs in the solar corona. Some other studies (see Cho et al., 2013) pointed out that CME-driven shocks can explain the source of type II radio bursts in the lower corona. It is widely recognized that type II radio bursts in the metric and deca-hectometric wavelengths associated with CMEs/ interplanetary coronal mass ejections (ICMEs) can be deployed as substitutes to examine the near-Sun kinematics and dynamics of CMEs and can serve as an indicator for geomagnetic storms (Mujiber et al., 2012; Vasanth and Umapathy, 2013; Shanmugaraju et al., 2017; Umuhire et al., 2021b).

In practice, the linear and quadratic fitting of the height-time data points gives the average speed and acceleration in the coronagraph/SDO field of view (FOV). Gopalswamy et al., (2013) provided that the type II height and the shock height at their respective onset times could be used as a proxy in studying the height-time profile of associated CMEs. Gopalswamy et al., (2012) derived the coronal magnetic field for June 13, 2010, type II burst in a 1.5 to 1.3 G range over a heliospheric distance of 1.3-1.5 Rs and concluded that EUV imagers coupled with radio dynamic spectra can be used as coronal magnetometers. Zimovets and Sadykov (2015) analyzed the spatial properties of February 16, 2011, type II radio burst from three different ground stations, including the Compound Astronomical Low frequency Low cost Instrument for Spectroscopy and Transportable Observatory (CALLISTO). The study concluded that the different emission tracks of the type II radio burst can be consequences of a single CME with different parts propagating through regions with different physical conditions, i.e., electron densities and geometries.

It is worth mentioning that spectral features of type II bursts can exhibit varying degrees of complexity. Such complexities in the harmonic or fundamental bands can be due to CME-CME interaction in the lower corona (see Lata Soni et al., 2021). This activity further alters the kinematics and magnetic field structures of other CME-associated events (Zhou and Feng, 2021). Recently, Umuhire et al., (2021b) statistically analyzed the properties of 51 high frequency type II bursts from CALLISTO spectrometer between 2010 and 2019 to explore the trends and emergence of the high frequency type II bursts. Conclusions from the study confirmed previous results that type II radio bursts are associated with wide and fast CMEs near the solar disk.

Since type II radio bursts originating from shocks observed in the solar corona and interplanetary medium (IPM) can give clues to space weather monitoring and forecasting; the present study aimed to statistically analyze the behavior of spectral properties of band splitting type II solar radio bursts and their relation with sunspot number in solar cycle 24. It is worth knowing that earlier studies already validated the CALLISTO spectrometer's capacity to provide real-time radio bursts data. Hence, it is justified to deploy and study only type II burst with split in frequency bands from ground stations within the e-CALLISTO network to assess the temporal variability of the physical conditions of these events during solar cycle 24. Parameters such as shock height, drift rate, bandwidth, Mach number, density jump, Alfvén speed, shock speed, magnetic field, and electron density were analytically extracted from the dynamic spectra to ascertain their relation with CMEs, sunspot number, and solar flares.

This article is structured as follows. Section 2 describes observations of solar radio data, CME, and solar flare events. Presented in section 3 are the analytical methods that were applied to extract the shock parameters from the type II radio bursts. The results and discussions are covered in section 4. The summary and conclusions of the study are provided in section 5.

## 2. Event observations and review.

*2.1 Type II radio bursts.*

The study implements statistical analyses of 20 type II radio bursts with band splitting sourced from the e-CALLISTO Spectrometer Network[1] to investigate their behavior with sunspot number, CMEs, and solar flares. CALLISTO is designed to observe solar radio transient emissions daily in a frequency range of 10 MHz - 870 MHz. Its frequency range can explore the solar corona in a heliospheric range of 1-3 Rs (solar radii). A detailed account of the CALLISTO spectrometer is presented by Benz et al., (2005, 2009); Pohjolainen et al., (2007); Zucca et al., (2012); Sasikumar Raja et al., (2018). The sample of events was observed for seven years (2010-2017) in solar cycle 24. The first step was a concurrent review of the National Oceanic and Atmospheric Administration (NOAA) solar and geophysical daily reports[2] and the data table presented by Umuhire et al., (2021b). The link to the quick views[3] of the radio bursts station catalog was followed to manually download FITS files of the selected radio bursts within the frequency range 20 MHz - 200 MHz based on the following criteria; 1) The type II radio burst should have band splitting in either the harmonic, fundamental, or both. 2) Assume band splitting in cases where there are complexities in type II structures. 3) The type II must be associated with CME and/or solar flare events. It is worth noting that the occurrence rate of type II bursts with splits in frequency bands is generally very low compared to their single-band counterparts. Fig.1 shows the geographical locations of seven CALLISTO ground stations within the e- CALLISTO network that collectively detected the solar radio bursts utilized in this present study.

The left panel of Fig.2 shows a dynamic spectrum of a low starting frequency type II sourced from DARO CALLISTO station-Germany in the e-CALLISTO network[4]. The event occurred on May 2, 2013, between 05:06 UT and 05:22 UT (see Nedal et al., 2019; Lata Soni et al., 2021). It was denoised by removing the background to mitigate radio frequency interferences (RFIs) to enhance the spectrum. The burst showed band splitting in the fundamental band. However, the harmonic band is relatively not clear. The type II radio burst is associated with CME having a mean linear speed of 671 km/s. There was also a preceding group of type III associated with M1.1 flare located at N10W25. The solar flare commenced at 04:58 UT and ended at 05:30 UT.

---

[1] http://www.e-callisto.org/
[2] ftp://ftp.ngdc.noaa.gov/STP/swpc_products/
[3] http://soleil.i4ds.ch/solarradio/callistoQuicklooks/
[4] http://soleil.i4ds.ch/solarradio/qkl/2013/05/02/DARO_20130502_050001_58.fit.gz.png

Similarly, the radio dynamic spectrum in Fig.2 (right panel) was observed by the OOTY CALLISTO station-India[5]. The event occurred on June 13, 2010, between 05:30 UT and 05:45 UT. It was also recorded by the Hiraiso Radio Spectrograph[6]. The fundamental and harmonic components of the radio burst are relatively visible. Thus, the fundamental component is less noticeable than the harmonic component. The band splitting in the harmonic band indicates that emission originates behind and ahead of the shock (see Smerd et al., 1975; Vršnak et al., 2001b; Cho et al., 2007; Hariharan et al., 2014). Its associated CME has an average speed of 320 km/s. The type II radio burst is also associated with M1.0 solar flare located at S23W75. The peak time of the solar flare is 05:39 UT.

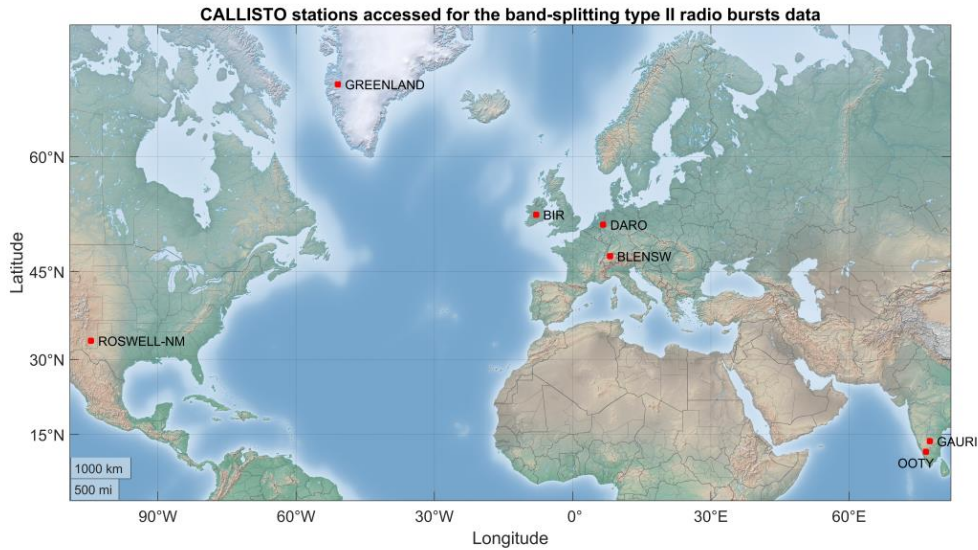

Fig. 1. Map showing ground stations within the e-CALLISTO network that detected the band splitting type II radio bursts considered in this study. Greenland station is located at Kangerlussuaq, Greenland, and operates at 10- 105 MHz. DARO is located at Hamminkeln, Germany, and works in 20- 80 MHz. GAURI and OOTY are stationed at Gauribidanur and Ooty, Tamil Nadu, India, working between 45- 410 MHz and 45- 445 MHz, respectively. BLENSW station is in Bleien, Switzerland, and functions at 150- 870 MHz. The BIR station is located at Birr Castle, Ireland, and operates within 20- 90 MHz. ROSWELL-NM spectrometer is stationed in New Mexico, USA and the working frequency range is 20- 90 MHz.

---

[5] http://soleil.i4ds.ch/solarradio/qkl/2010/06/13/OOTY_20100613_053001_58.fit.gz.png
[6] https://sunbase.nict.go.jp/solar/denpa/hirasDB/Events/2010/2010061305.jpg

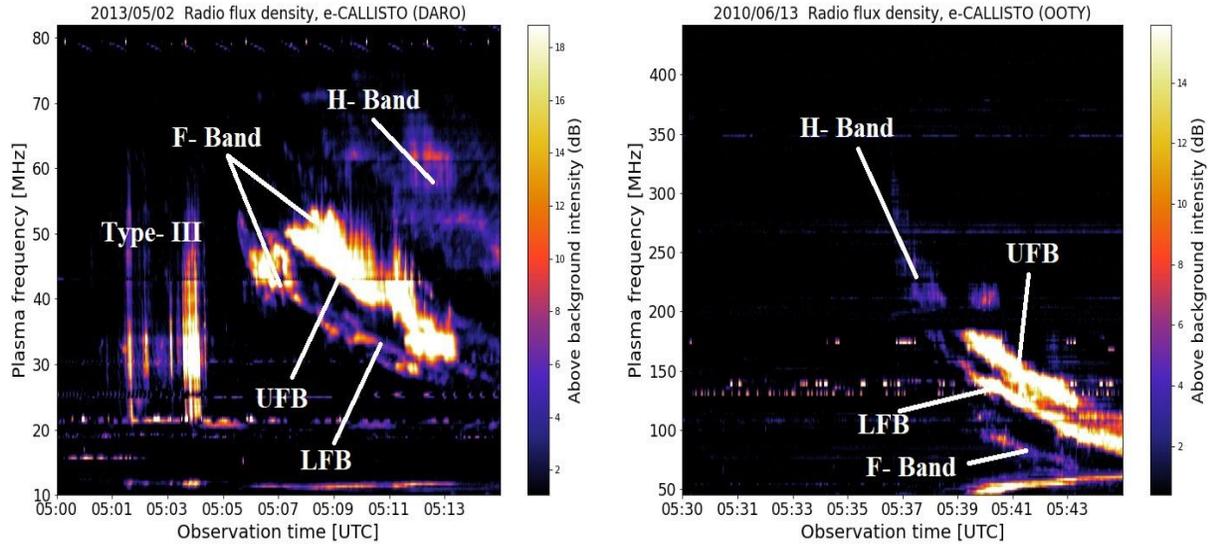

Fig. 2. (left) Radio dynamic spectrum observed from DARO-Germany at 05:06 UT on May 05, 2013. The harmonic band is barely visible whiles the fundamental band dominates, with band splitting having patches in the lower frequency band. There was a preceding group of type III between 05:02 UT and 05:05 UT. (Right) Type II radio dynamic spectrum was observed from OOTY-India on Jun 13, 2010, at 05:30 UT. The fundamental band is barely visible, and the harmonic band is the dominant feature with band splitting. The event was associated with M1.0 solar flare located at S23W75.

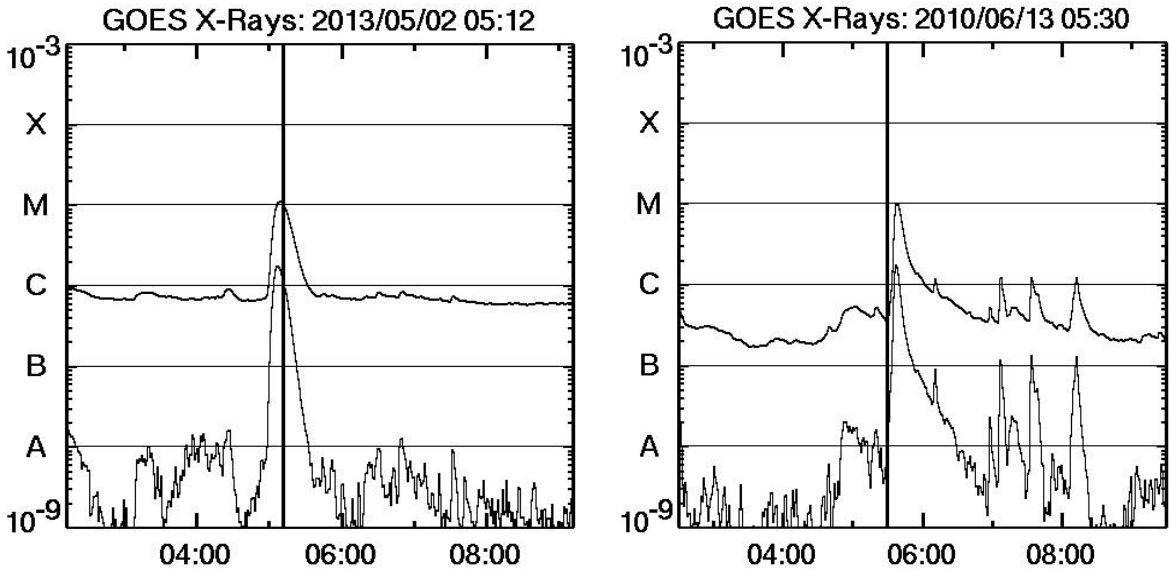

Fig. 3. GOES X-ray light curve for solar flare categorization for CME associated solar flares. (Left) represents the CME associated solar flare taken at the peak time (05:12 UT) on May 02, 2013. (Right) represents the solar flare classification curve window taken on June 13, 2010, at 05:30 UT. The vertical lines in both windows show the eruption times from AIA images of the associated CMEs that co-related with the peaks of the associated flares.

Fig.3 shows the solar flare classifications by the GEOS X-ray light curves for the two radio events under review. The upper and lower curves in the GOES soft X-ray plot are given by 1-8 Å and 0.4-5 Å channels, respectively. In each case, the peak of the curves denotes the flare classes and their corresponding peak times. The vertical lines represent the eruption times of the associated CMEs that are nearly co-related with the peaks of the associated flares.

*2.2 CME and solar flare observations.*

An extensive survey was conducted by utilizing the Coordinated Data Analysis Workshop (CDAW)[7] online CME database and the Heliospheric Event Registry[8] simultaneously to check whether the type II radio bursts were associated with CMEs and solar flares. The linear speed and the estimated time of the associated CMEs were obtained from the SOHO/LASCO CME catalog[9]. The catalog also directs the Geostationary Operational Environmental Satellite (GOES) to monitor the flare proprieties for CMEs that correspond to the designated type II bursts under review. Hence, the peak times, locations, and the classes of the associated flares were compiled.

**3. Method.**

*3.1. Estimating the physical properties of type II radio bursts.*

Type II radio bursts are generated by propagating shocks that form electron beams, causing the emission of Langmuir waves. This process is based on the plasma emissions around fundamental and harmonic frequencies. The relationship between the plasma frequency $f_p$ (MHz) and electron density n($cm^{-3}$) in the expression below was used to ascertain information about the plasma density at which the emission occurs (see Smerd et al., 1975).

$$f_p = 9 \times 10^3 \sqrt{n} \quad (1)$$

One possible explanation is that the frequency drift towards lower frequencies indicates a decrease in the magnitude of electron density. This variation reiterates that solar radio burst drivers are progressing towards higher heights and lower densities in the solar atmosphere. The region ahead (upstream) and behind (downstream) the shock emission have relative magnitudes in terms of their respective frequencies ($f_1$, $f_2$) and electron densities ($n_1$, $n_2$). This implies that the relative instantaneous bandwidth (BDW) of the emission lines can be computed as (see Vrˇsnak et al., 2001b),

$$BDW = \frac{\Delta f}{f} = \frac{f_2 - f_1}{f_1} = \sqrt{\frac{n_2}{n_1}} - 1 \quad (2)$$

Applying the instantaneous bandwidth (BDW) equation, the density jump (X) across the shock can be estimated as (see Vrˇsnak et al., 2001b),

$$X = \frac{n_2}{n_1} = (BDW + 1)^2 \quad (3)$$

Assuming a shock is perpendicular to the magnetic field and plasma $\beta \to 0$, then the density jump equation coupled with the simplified Rankine-Hugonoit equation can be used to estimate the Alfvén Mach number ($M_A$),

---

$$M_A = \sqrt{\frac{X(X+5)}{2(4-X)}} \qquad (4)$$

Fig.4 (a) shows a comparison between density jump and Mach number as functions of shock height. The drift rate was estimated to explore the CME-driven shock temporal evolution of the type II bursts under consideration using the equation,

$$D[MHz/s] = \frac{\Delta f[MHz]}{\Delta t[s]} = \frac{f_2 - f_1}{t_2 - t_1} \qquad (5)$$

where D is the drift rate, $f_1$, $f_2$ are the starting and ending frequencies, respectively and $t_1$, $t_2$ are the corresponding times. The above equation is related to the shock speed and the density gradient in the solar corona (Gopalswamy, 2011). The graphical representation of the drift rate and the shock speed against time is shown in Fig.4 (b). The drift rate of the fundamental band, the emission frequency, and the Newkirk electron density model (Newkirk, 1961) were applied to estimate the height of the radio source of the type II emissions and the shock speed. The height-time profile of the lower-frequency band ($f_{LFB}$) and the upper-frequency band ($f_{UFB}$) is presented in Fig.4 (c). The estimated CME/shock speed ($V_S$) together with the Alfvén Mach number was utilized to estimate the Alfvén speed ($V_A$) using the relation,

$$V_A[km/s] = \frac{V_S}{M_A} \qquad (6)$$

The coronal magnetic field strength (B) in Gauss was determined from equation 7 using $V_A$ and emission frequency of the lower-frequency band ($f_{LFB}$) (see Vasanth et al., 2014). Fig.4 (d) shows a graphical illustration of the magnetic field strength (B) and drift rates versus shock height.

$$B[G] = 5.1 \times 10^{-5}(f_{LFB})(V_A) \qquad (7)$$

The standard error (SE) expressed in equation 8 was utilized as a metric for assessing the accuracy of the extracted parameters.

$$SE = \frac{\sigma}{\sqrt{n_m}} \qquad (8)$$

Where σ and $n_m$ represent the standard deviation and number of measurements, respectively. Note: In this section, the estimated properties of the type II radio burst event that occurred on Oct 25, 2013, at 13:30 UT were presented to illustrate the shock parameters graphically as a proxy for all the bursts under consideration.

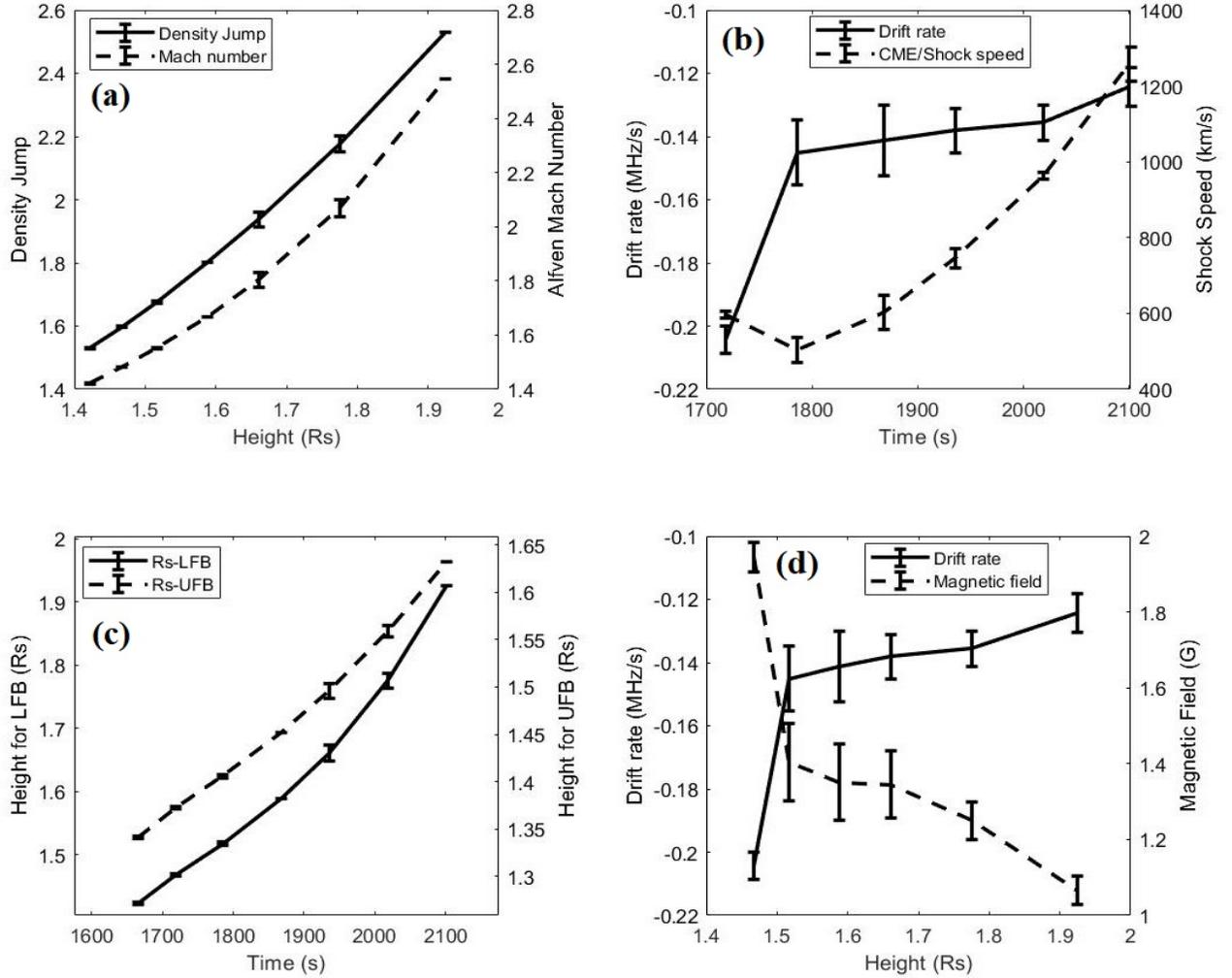

Fig. 4. Shock parameters extracted from type II radio burst observed on Oct 25, 2013, at 13:30 UT. (a) shows the plots of density jump and Mach number as functions of shock height (Rs). (b) represents plots of drift rates and CME/shock speed as functions of time. (c) illustrates the height-time profiles for the lower-frequency band (LFB) and upper-frequency band (UFB). (d) compares the behavior between drift rates and magnetic field as functions of shock height. The error bars were generated from the standard errors as metrics for evaluating the uncertainties.

## 4. Results and discussions.

*4.1 Statistical results.*

After a thorough survey of type II bursts with band splitting as described in section 2 of this study, it is noteworthy to reiterate that all the 20 type II radio bursts analyzed are associated with CMEs and solar flares. This is consistent with prior research by Gopalswamy et al., (2012, 2013), who concluded that CME-driven shocks could trigger type II radio bursts in the lower corona. Table 1 lists band splitting type II radio bursts and the associated CMEs/ solar flares from 2010-2017. Columns 1-11 in Table 1 list the type II parameters in the following sequence; date, time, lower frequency ($F_L$), upper frequency ($F_U$), shock height (R), bandwidth (BDW), CME/shock speed ($V_S$), Alfvén speed ($V_A$), Mach Number ($M_A$), density jump (X) and magnetic field strength (B). Columns 12-13 list the CME time and average speed from SOHO/LASCO. The flare peak time, location, and class are listed in columns 14-16. The last column (17) contains the weighted factor ($W_F$) regarding criteria 2, where 1 denotes 100% band splitting either in the fundamental or harmonic bands and 2 indicates events with complex frequency band structures with unclear band splitting. The yearly comparison between the average CME speeds from SOHO/LASCO and the estimated CME/shock speeds from the dynamic spectra is presented in Fig.5 (a). The black and grey bars represent the average CME/shock speeds from the type II radio bursts and CME average speeds from SOHO/LASCO, respectively. It can be deduced from the plot that the average linear speeds of the CMEs are slightly greater than CME/shock speeds estimated from the dynamic spectra in 2012, 2013, 2014, and a drastic margin in 2017, whiles the remaining years also have the reverse. However, this indicates that years with relatively low CME/shock speeds might have associated CMEs reaching their peak speeds before the start of the type II bursts, whereas the opposite indicates consistency with shock formation.

Fig.5 (b) shows the relationship between CME mean linear speeds within LASCO FOV and the average CME/shock speeds estimated from the dynamic spectra. The correlated coefficient (R) recorded was ~0.87 and ~0.93 after excluding the single extreme data point in red. The high correlation coefficient value is very significant and emphasizes that in the absence of white-light data, the speed estimated from radio data can also be deployed as a proxy for disc events to assess the initial kinematics and dynamics of a CME in the lower corona (see Kumari et al., 2017a, 2019; Umuhire et al., 2021a, 2021b). Fig.5 (c) shows a good correlation (R = 0.75) between the CME average linear speeds from SOHO/LASCO and the mean Alfvén speeds. Similarly, Fig.5 (d) presents a good correlation (R = 0.79) between the average drift rates and the mean plasma frequencies. Both results are significant, and the latter reiterates that the drift rates of metric type II bursts depend on the emission frequencies suggesting that type II radio bursts in the lower and middle corona (1-3 Rs) are excited by CMEs ahead and behind the shocks (see Vršnak et al., 2002; Gopalswamy et al., 2008). This result also showed good consistency with that obtained by Umuhire et al., (2021a), who analyzed 128 type II radio bursts with emphasis on 40 high-frequency type II bursts and found that the correlation coefficient between 88 low-frequency type II bursts and their drift rates stood at 0.74. Note that the large scatter in plots (c and d) is due to data values corresponding to different days within the solar cycle 24 and the daily variability in electron densities.

Table 1: Properties of band splitting type II radio bursts, CMEs, and solar flares.

| [1]Date (dd/mm/yy) | [2]Time$^{II}$ (UT) | [3]F$_L$ (MHz) | [4]F$_U$ (MHz) | [5]R(Rs)/ 4xN | [6]BDW | [7]V$_S$ (km/s) | [8]V$_A$ (km/s) | [9]M$_A$ | [10]X | [11]B(G) | [12]Time$^{CME}$ (UT) | [13]V$_{CME}$ (Km/s) | [14]Time$^{SF}$ (UT) | [15]Loc (deg) | [16]Class | [17]W$_F$ |
|---|---|---|---|---|---|---|---|---|---|---|---|---|---|---|---|---|
| 13/06/10 | 5:30:01 | 39.9 | 54.9 | 2.1 | 0.38 | 354 | 203 | 1.78 | 1.91 | 0.59 | 5:30:05 | 310 | 5:39:00 | S23W75 | M1.0 | 1 |
| 08/03/11 | 3:45:00 | 18.6 | 29.6 | 2.55 | 0.59 | 793 | 305 | 2.63 | 2.56 | 0.29 | 4:12:05 | 732 | 3:45:00 | S21W72 | M1/5 | 2 |
| 01/10/11 | 9:45:00 | 115.9 | 135.9 | 1.72 | 0.17 | 493 | 382 | 1.29 | 1.37 | 2.67 | 9:36:07 | 448 | 9:59:00 | N09W04 | M1.2 | 1 |
| 03/06/12 | 17:59:49 | 95.4 | 160.4 | 1.84 | 0.68 | 558 | 185 | 3.09 | 2.83 | 1.53 | 18:12:05 | 605 | 17:55:00 | N16E38 | M3.3 | 2 |
| 13/0113 | 8:29:58 | 89.8 | 112.8 | 1.88 | 0.25 | 528 | 362 | 1.46 | 1.58 | 2.1 | 8:48:05 | 696 | 8:23:00 | S20E89 | B4.1 | 1 |
| 02/05/13 | 5:00:01 | 53.8 | 88.8 | 1.76 | 0.65 | 571 | 202 | 2.94 | 2.74 | 0.56 | 5:24:05 | 671 | 5:10:00 | N10W25 | M1.1 | 1 |
| 25/10/13 | 13:30:00 | 59.6 | 80.6 | 2.29 | 0.37 | 806 | 449 | 1.76 | 1.89 | 1.83 | 15:12:09 | 1081 | 13:37:00 | N06W24 | C2.3 | 1 |
| 08/11/13 | 4:14:59 | 85 | 135 | 1.52 | 0.6 | 526 | 201 | 2.56 | 2.53 | 0.86 | 3:24:07 | 497 | 4:26:00 | S13E13 | X1.1 | 2 |
| 07/12/13 | 7:16:54 | 27.4 | 47.4 | 2.34 | 0.74 | 1083 | 316 | 3.87 | 3.06 | 0.47 | 7:36:05 | 1085 | 7:29:00 | S16W48 | M1.2 | 2 |
| 08/01/14 | 3:45:01 | 43.6 | 56.6 | 2.82 | 0.3 | 555 | 355 | 1.57 | 1.69 | 1.03 | 4:12:05 | 643 | 3:47:00 | N11W88 | M3.6 | 1 |
| 20/02/14 | 7:29:58 | 44.1 | 62.1 | 2 | 0.41 | 954 | 557 | 1.85 | 1.98 | 1.28 | 8:00:07 | 948 | 7:56:00 | S15W75 | M3.0 | 2 |
| 22/08/14 | 10:15:00 | 115.2 | 154.2 | 1.45 | 0.34 | 196 | 119 | 1.67 | 1.8 | 0.71 | 10:00:05 | 217 | 10:27:00 | N11E01 | C2.2 | 1 |
| 05/11/14 | 9:44:57 | 32.4 | 42.4 | 3.38 | 0.31 | 400 | 252 | 1.59 | 1.72 | 0.55 | 10:00:05 | 386 | 9:47:00 | N20E68 | M7.9 | 1 |
| 11/03/15 | 16:15:00 | 82.9 | 102.9 | 1.94 | 0.24 | 247 | 175 | 1.43 | 1.54 | 0.53 | 17:00:05 | 240 | 16:22:00 | S17W21 | X2.1 | 1 |
| 01/06/15 | 13:30:00 | 147.9 | 181.9 | 1.3 | 0.23 | 787 | 345 | 1.41 | 1.51 | 2.68 | 13:48:05 | 748 | 13:30:00 | N06W17 | C1.4 | 1 |
| 22/08/15 | 6:45:00 | 82.7 | 121.7 | 1.6 | 0.47 | 527 | 252 | 2.06 | 2.17 | 1.04 | 7:12:04 | 547 | 6:49:00 | S14E09 | M1.2 | 1 |
| 02/05/16 | 8:30:00 | 106.8 | 143.8 | 1.77 | 0.35 | 175 | 106 | 1.7 | 1.83 | 0.8 | 9:12:09 | 262 | 8:42:00 | N21E33 | C3.5 | 1 |
| 10/07/16 | 1:00:00 | 92.6 | 137.6 | 1.4 | 0.49 | 547 | 251 | 2.13 | 2.22 | 1.16 | 1:25:44 | 368 | 0:59:00 | N12E67 | C8.6 | 2 |
| 02/09/17 | 15:39:58 | 59.7 | 83.7 | 1.79 | 0.4 | 636 | 349 | 1.84 | 1.97 | 1.07 | 15:48:05 | 705 | 15:37:00 | N14E45 | C2.7 | 2 |
| 06/09/17 | 12:00:00 | 66.6 | 97.6 | 2.12 | 0.46 | 900 | 453 | 2.04 | 2.15 | 2.3 | 12:24:05 | 1571 | 12:02:00 | S09W34 | X9.3 | 2 |

[2]Time$^{II}$ = type II onset time, [12]Time$^{CME}$ = CME Time from SOHO/LASCO, [14]Time$^{SF}$ = solar flare peak time, [15]Loc = location of solar flare
[3]F$_L$, = average lower frequency, [4]F$_U$ = average upper frequency (either fundamental or harmonic)
[5]R = estimated shock height (solar radii), [6]BDW = Bandwidth, [7]V$_S$ = estimated CME/shock speed from type II radio bursts, [8]V$_A$ = Alfvén speed
[9]M$_A$ = Mach number, [10]X = density jump, [11]B = magnetic field strength, [13]V$_{CME}$ = CME speed from SOHO/LASCO
[17]W$_F$ = weighted factor where 1 denotes 100% certainty that band splitting is present either in the fundamental or harmonic band and 2 denotes uncertainty of band splitting due to complexities in the structures of those type II bursts.

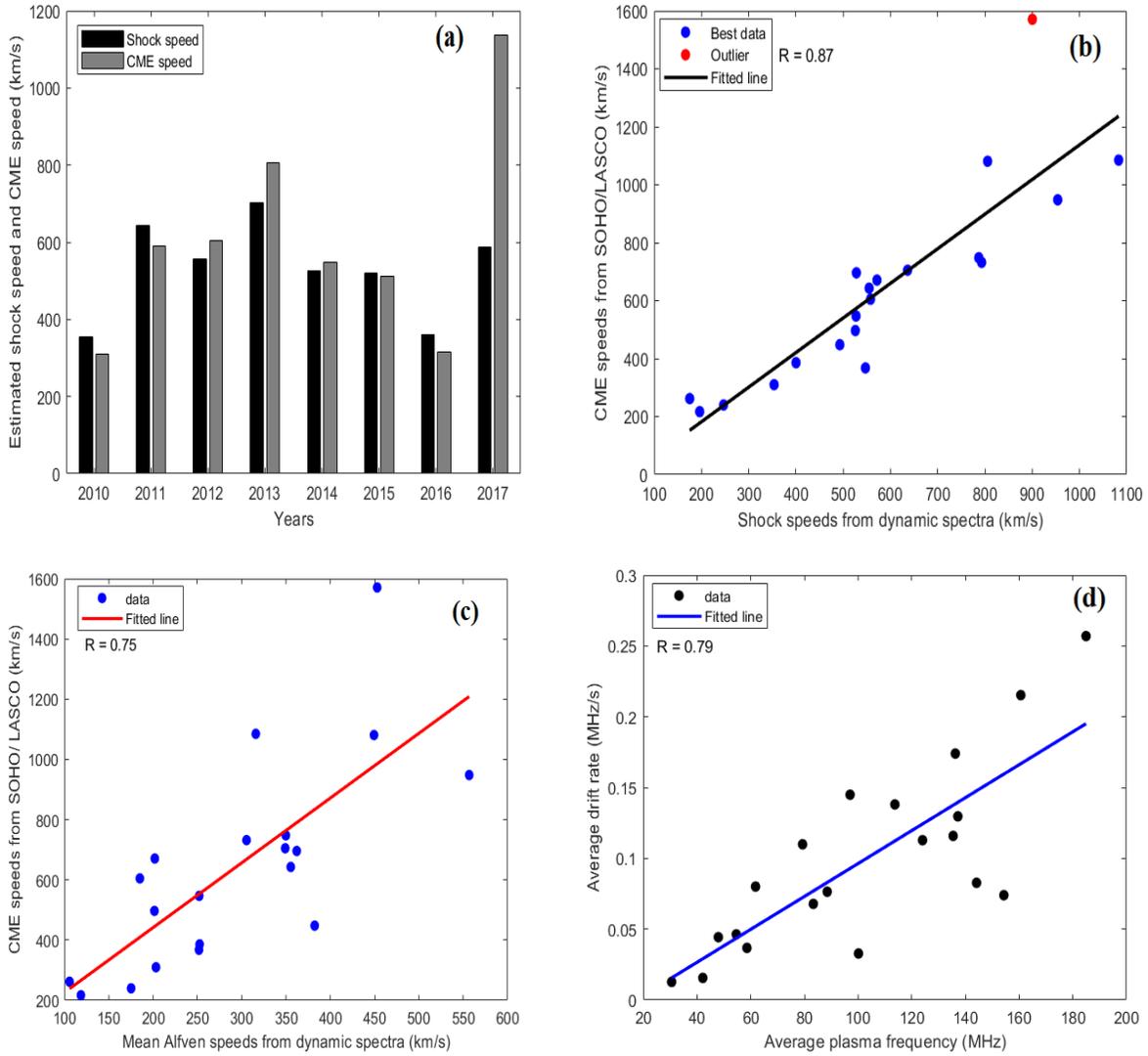

Fig. 5. (a) Comparison between the yearly variability of average shock speeds estimated from the band splitting type II bursts observed by several ground stations within the e-CALLISTO network from 2010 to 2017 and the associated CME linear speeds observed in SOHO/LASCO represented in black and grey, respectively. (b) The correlation between the SOHO/LASCO CME linear speed and the estimated CME/shock speeds (mean). (c) Relationship between the CME linear speed and average Alfvén speed estimated from the dynamic spectra. (d) The correlation between absolute mean drift rate and mean plasma frequency. The correlation coefficients are 0.87, 0.75, 0.79 respectively. However, the correlation coefficient obtained in plot b improves to ~0.93 after excluding the outlier.

A comparison between the estimated CME/shock speeds and the shock heights with the source longitudes and latitudes (i.e., solar flare locations) is presented in Fig.6. It can be inferred from Fig.6 (a) that the majority of the flare-associated type II radio bursts within the southern and northern hemispheres have CME/shock speeds between 200 km/s and 800 km/s. Only two type II events have extreme shock speeds ($>1000 km/s$). The red and blue colors represent the southern and northern hemispheres, respectively. One plausible justification for these findings is that the increase in variability of type II CME/shock speeds within the northern and southern hemispheres can be due to the increase in coronal holes within these regions before the onset of high speed solar wind during the solar cycle 24 (see Nakagawa et al., 2019). Fig.6 (b) illustrates the variation of the estimated CME/shock speed of the type II radio bursts with the heliographic longitudes. It is noticeable that more type II bursts emanate from western longitudes with much higher CME/shock speeds and variability than type II bursts originating from the eastern longitudes. The red and blue colors indicate the eastern and western longitudes, respectively. This result established that the type II associated CMEs within the western longitudes have enough speeds to drive a shock. Also, a more significant number of type II radio bursts within the region are due to shock waves influencing the directivity of type II radio bursts (see Ramesh et al., 2013; Gopalswamy et al., 2016; Mahender et al., 2020). The result also suggests that type II solar radio bursts originating from the western longitudes may generate significant Solar Energetic Particle (SEP) fluxes.

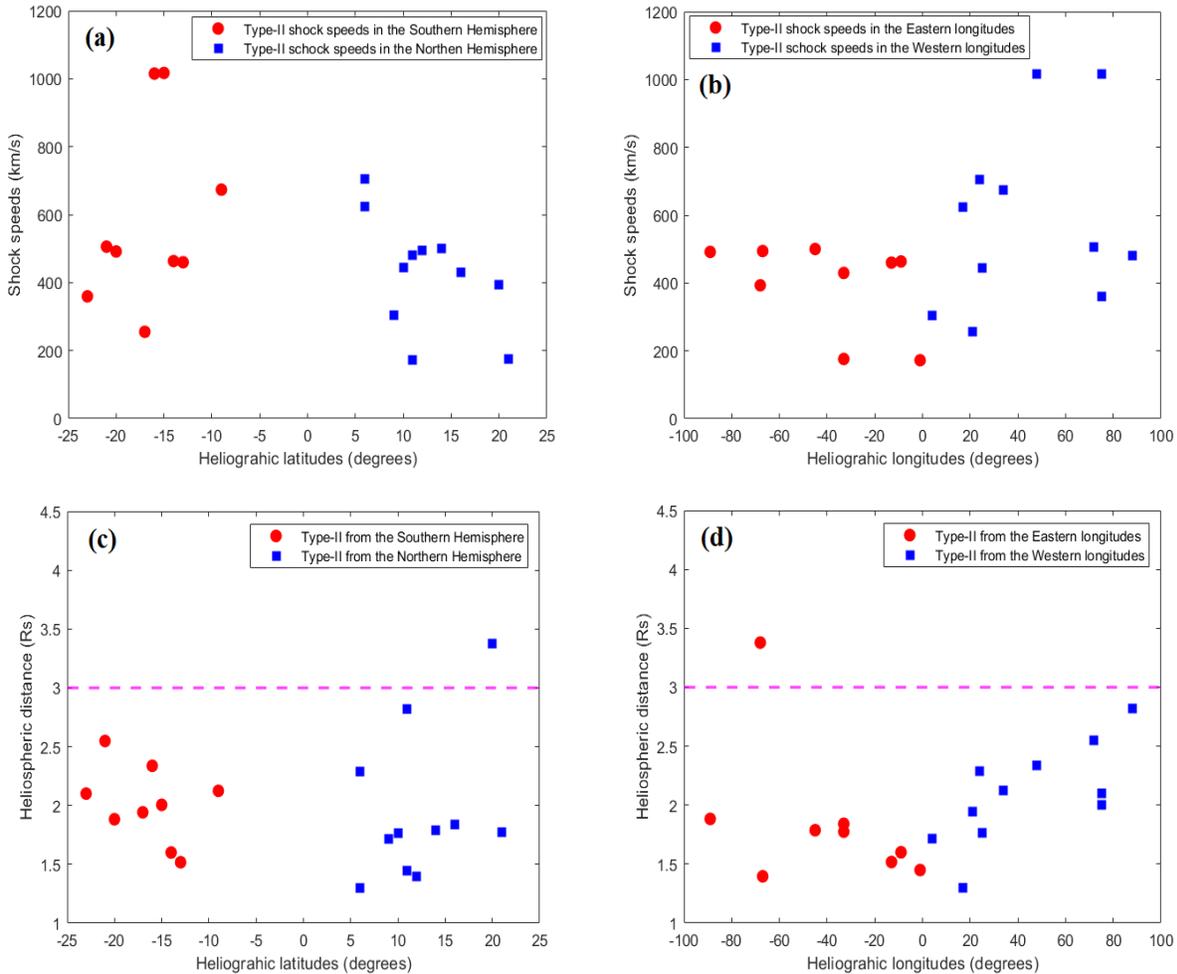

Fig. 6. Relationship between CME/shock speeds and shock heights of type II radio bursts with heliographic latitudes and longitudes. Panel (a) shows the variation of type II shock speed with heliographic latitudes, and panel b shows type II shock speed with heliographic longitudes. Similarly, panel (c) shows the variation of type II shock heights with heliographic latitudes, and panel d shows type II shock heights with heliographic latitudes. The reference line (magenta) shows the maximum heliospheric distance (~3) according to the Newkirk coronal density model.

Equivalently, Fig.6 (c and d) illustrates the variation of average shock heights of the type II radio bursts with the heliographic latitudes and longitudes of the associated solar flares. Most of the analyzed type II bursts are found to be within the acceptable heliospheric height range (1-3 Rs). It must be emphasized that these results are not entirely different from the preceding analyses on the shock speed longitudinal-latitudinal variability. However, one type II event with a shock formation height greater than 3 Rs was observed. This can be explained that the shock might have been sparked at CME flanks or experienced flare-blast waves (see Shanmugaraju et al., 2017).

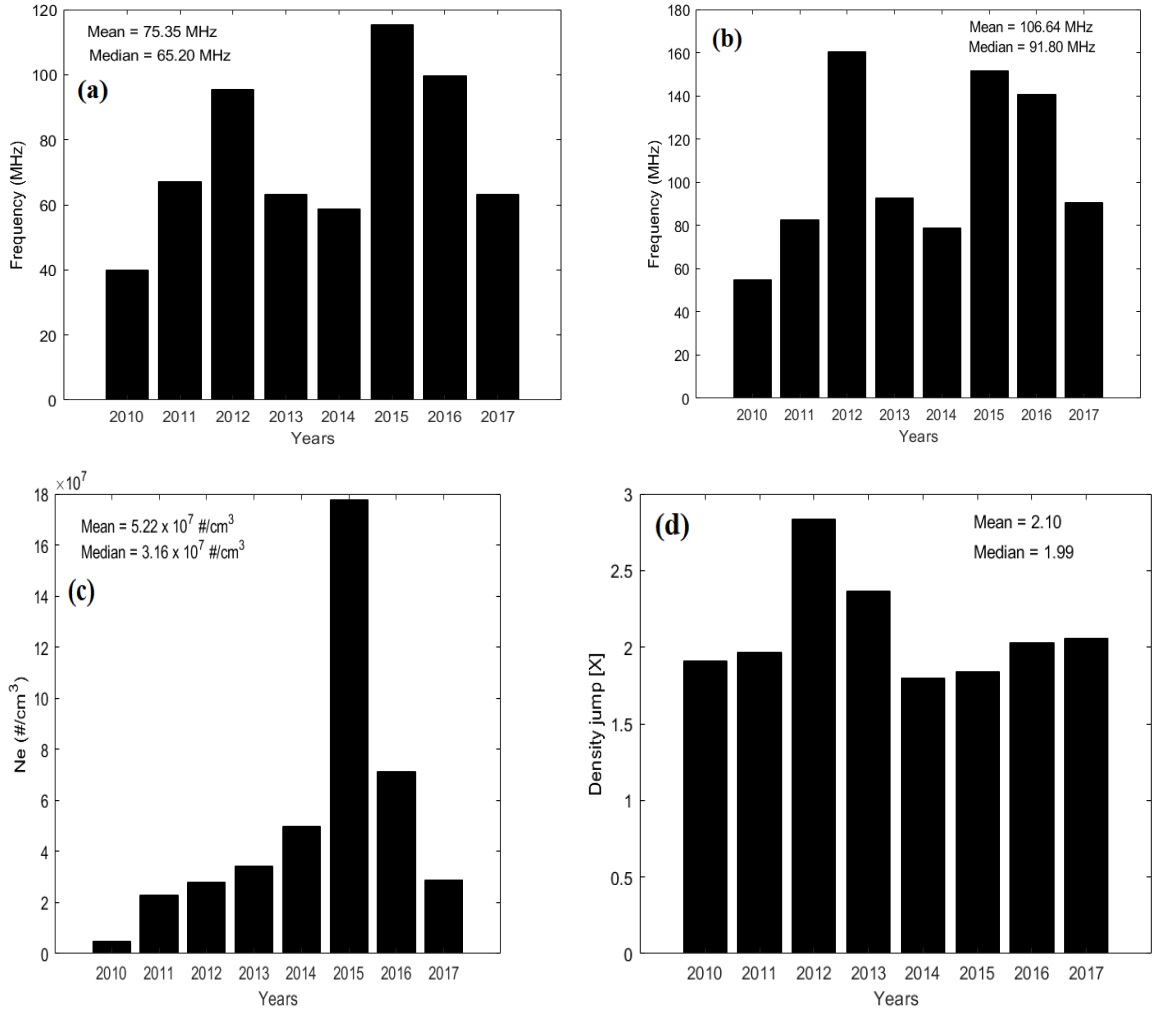

Fig. 7. Yearly distribution of (a) Lower frequency (b) Upper frequency (c) Electron density (d) Density jump estimated from type II dynamic spectra using coronal Equations 1, 3, and the 4-fold Newkirk electron density model.

Fig.7 presents the yearly distribution of the average lower-frequency, average upper-frequency, electron density, and density jump estimated from the radio dynamic spectra. It is observed that the distribution in Fig.7 (a) is roughly bimodal, with mean and median values of 75.35 MHz and 66.20 MHz, respectively. Band splitting type II events observed in 2010 and 2017 have occurred at relatively low frequencies, while events in 2015 have the highest frequencies. This yearly frequency variation is a significant signature indicating that the analyzed type II radio bursts can be correlated with solar cycle 24. Similarly, the annual variation of the upper frequency given in Fig.7 (b) depicts nearly the same variations of lower frequencies in panel a. The mean and median values are 106.64 MHz and 91.80 MHz, respectively. These values show a degree of consistency with results obtained by Gopalswamy et al., (2013), who analyzed 32 metric type II bursts and found that the mean and median starting frequencies are 102 MHz and 85 MHz, respectively. The yearly variation of the coronal electron density is presented in Fig.7 (c). There was a gradual increase in electron density from 2010 to 2014 and an abrupt increase in 2015, with a steady decline from 2016 to 2017. The mean and median are $5.22 \times 10^7 Ne/cm^3$ and $3.16 \times 10^7 Ne/cm^3$. This variation is significant to examine the near-Sun coronal dynamics in solar cycle 24. Panel (d) of Fig.7 shows the yearly distribution of the density jump. The peak of the density variation was recorded in 2012. The distribution is quite symmetrical, with a mean and median of 2.10 and 1.99, respectively.

In an attempt to answer questions relating to the physical conditions of type II radio bursts during solar cycle 24, a different dimension was considered to analyze the variability and the synchronization of the spectral properties of type II bursts with sunspot number. These analyses presented in Fig.8 are comparatively different from previous works by some authors who attempted to analyze the occurrence rate (number) of solar radio bursts, primarily type III radio bursts (Lobzin et al., 2011; Huang et al., 2018; Mahender et al. 2020; Ndacyayisenga et al., 2020; Morosan et al., 2021). For instance, Ndacyayisenga et al., (2020) analyzed the occurrence rate of 12971 type III bursts spanning 2010 to 2017. It is evident from their study that the yearly distribution of the number of type III bursts strongly correlates with sunspot number during solar cycle 24.

However, Fig.8 of this current study presents the association between the yearly variability of the type II shock heights, average magnetic field, mean Alfvén speed, average CME/shock speed, and the revised version of the sunspots number (magenta curve)[10] (Clette et al., 2016). In Fig. 8 (a), it can be visually deduced that the shock heights of the type II bursts correlate with the monthly averaged smoothed sunspot number by following the major peak (2014) of the sunspot number that occurred during the solar cycle 24. Thus, the shock heights of the type II bursts roughly synchronize with solar cycle 24 variations. The distribution is quite symmetrical, with mean and median values of 1.97 and 1.86, respectively. Fig.8 (b) shows the annual magnetic field strength variations with sunspot number. It can be inferred that the magnetic field strength of the type II bursts significantly correlates with the monthly averaged smoothed sunspot number by tracking the trajectory of the ascending and declining phases of solar cycle 24, with the minor peak coinciding with the magnetic field strength in 2012. Contrary, the magnetic field strength in 2014 did not synchronize with the major peak of the solar cycle 24. One possible explanation is that at higher radial distances (Rs), the magnetic field lines are open (low magnetic field strength), and such type II events occurring in the outer solar corona have lower drift rates and lower frequencies. The average and median of the magnetic field are 1.0 G and 0.8 G, respectively. This result is in good agreement with the ranges reported in previous studies (see Gopalswamy et al., 2012; Cunha-Silva et al., 2014; Vasanth et al., 2014; Kishore et al., 2016; Lata Soni et al., 2021).

---

[10] https://wwwbis.sidc.be/silso/datafiles

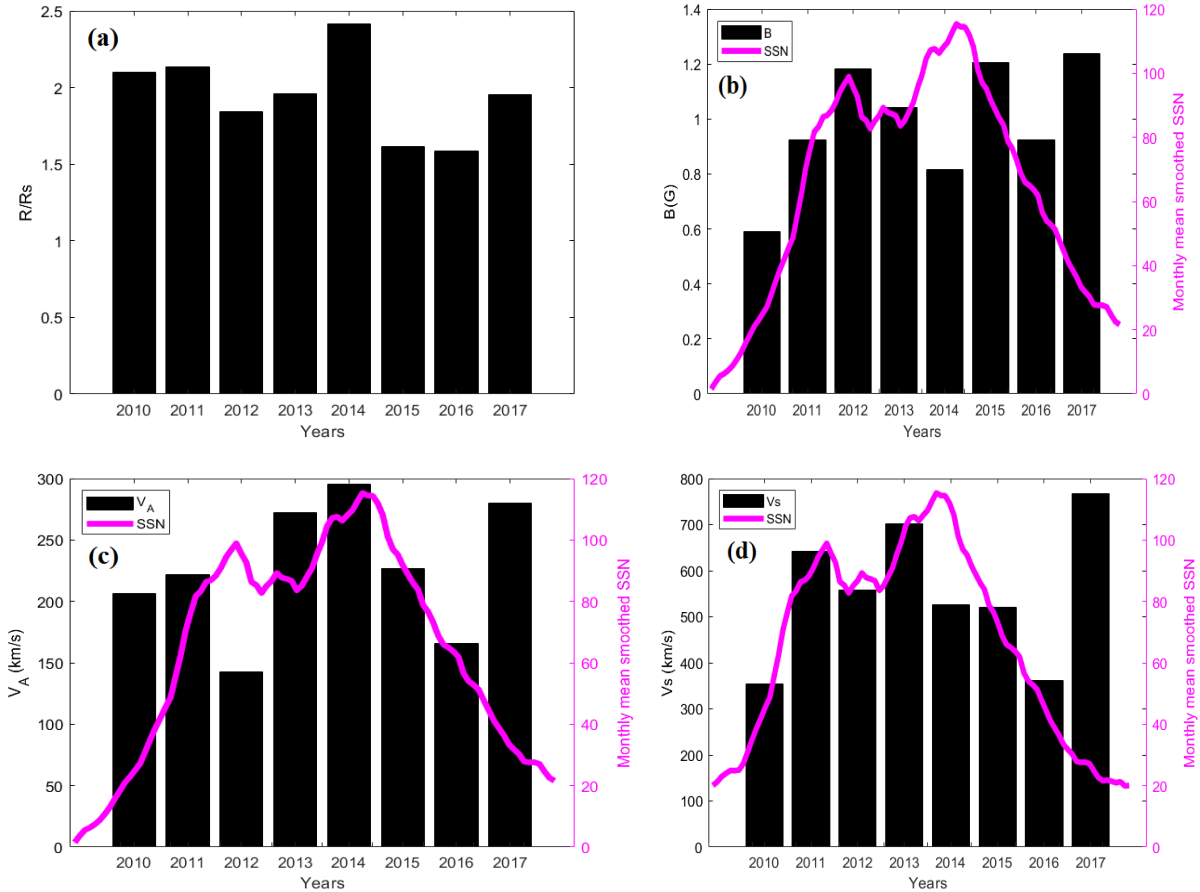

Fig. 8: Distribution of type II bursts physical parameters; (a) Rs (solar radii), (b) magnetic field strength (G), (c) Alfvén speed (km/s), (d) shock speed (km/s)] and their association with solar activity. The magenta curve represents the monthly averaged sunspot number.

In Fig.8 (c), it is observed that the Alfvén speed also followed the trajectories of the ascending and declining phases, with the maximum speed synchronizing with the major peak of the solar cycle 24. It is important to note that type II radio bursts recorded in 2012 and 2016 have relatively lower Alfvén speeds. The mean and median of Alfvén speed are 247 km/s and 232 km/s, respectively and were found to be within the Alfvén speed range reported by (Vršnak et al., 2001a; Gopalswamy et al., 2012). The shock speed variations with sunspot number are presented in Fig.8 (d). Inference from the plot indicates that the shock speed significantly correlates with the sunspot number by roughly tracing the path of the solar cycle 24 during the ascending and declining phases. The mean and median of the CME/shock speeds estimated from the low-frequency type II radio bursts provided by CALLISTO ground stations are 582 km/s and 551 km/s, respectively. It is noteworthy that this result is consistent with previous studies conducted by Ramesh et al., (2010); Cunha-Silva et al., (2014), who analyzed type II radio bursts sourced from CALLISTO (Brazil) and Gauribidanur radioheliograph (GRH, India) stations and reported that the CME/shock speeds are in the range of 452-1389 km/s and 487-974 km/s respectively.

**5. Summary and conclusion.**

This study conducted detailed investigations into the spectral properties of 20 low-frequency type II radio bursts with band splitting and their association with CMEs, solar flares, and sunspot number. The type II events were sourced from seven CALLISTO spectrometers during solar cycle 24 (2010 -2017). The list of events was based on the selection criteria stated in section 2.

Statistical analysis was conducted to explore the temporal variability between the estimated CME/shock speeds and the averaged CME speeds from the SOHO/LASCO FOV. It was observed that estimated CME/shock speeds from the dynamic spectra were higher than CME speeds from the SOHO/LASCO database in some years within the solar cycle 24, whiles the remaining years hold the reverse. This confirms previous studies that years with relatively low CME/shock speeds might have CMEs reaching their peak speed before the onset of the type II bursts, while the opposite is true for type II bursts originating ahead of the CMEs. Also, the estimated CME/speeds from the dynamic spectra were compared with the associated CME speeds from SOHO/LASCO FOV to assess and validate the performance of the coronal density model deployed in this study. After neglecting the outlier, the correlation coefficient (R= 0.87) improved to 0.93.

The Alfvén speeds from the radio dynamic spectra and CME speeds from SOHO/LASO FOV showed a significant correlation of $R = 0.75$, and that between the absolute drift rates and plasma frequencies was 0.79. The latter suggests that type II solar radio bursts during solar cycle 24 are excited by CMEs ahead and behind the shocks. It is worth mentioning that most flare-associated type II bursts originate within the northwest region and have average CME/shock speeds between 200 km/s and 800 km/s. Furthermore, the studies revealed that the number of solar flare-associated type II radio bursts from the eastern longitudes is relatively lesser than those from the western longitudes. It is noteworthy that since the variability of the type II bursts at the flare sites is relatively uniform, this indicates that the type II radio bursts are non-directive (see Singh et al., 2019).

Finally, the study assessed the temporal variability of the estimated shock parameters from the dynamic spectra and observed that; (1) the lower and upper frequency of all type II events exhibited a bimodal distribution during solar cycle 24; (2) the electron density is lowest at the start of the cycle (i.e., 2010) and peaks at the beginning of the declining phase of solar cycle 24 (i.e., 2015); (3) the peak of the density jump recorded in 2012 synchronizes with the minor peak of the solar cycle 24. Additionally, the study ascertained that the temporal distribution of the shock heights, magnetic field strengths, CME/shock speeds, and Alfvén speeds estimated using the coronal density model correlates with the trajectory of the solar cycle 24. The present study emphasizes the uniqueness of band splitting type II solar radio bursts and their spectral properties as potential signatures in monitoring and predicting space weather hazards. The study further demonstrates the significance of deploying more CALLISTO ground stations for space weather monitoring and forecasting.

**Acknowledgments.**

This research was financially supported by the EJUST-TICAD7 scholarship, a partnership program between the government of Japan and the Egyptian government. We are grateful to the e-CALLISTO data center hosted by FHNW, Institute for Data Science-Switzerland and managed by Christian Monstein, the principal inventor of CALLISTO; NOAA-SWPC; GOES; SOHO/LASCO; CDAW and WDC- SILSO, Royal Observatory of Belgium, Brussels, for their policies on open data accessibility. The author (Felix N. Minta) registers his heartfelt gratitude to Mohammed Nedal (Institute of Astronomy with National Astronomical Observatory, Bulgarian

Academy of Sciences, Sofia, Bulgaria) for his immense support in providing part of the algorithms used in this project. Our unqualified appreciation goes to the anonymous neutral reviewers for their valuable and encouraging comments in improving this manuscript.

**References**

Benz, A. O., Monstein, C., Meyer, H. *et al.,* 2009. A world-wide net of solar radio spectrometers: e-CALLISTO. Earth, Moon and Planet. 104(1), 277–285. https://doi.org/10.1007/s11038-008-9267-6

Benz, A. O., Monstein, C., Meyer, H., 2005. CALLISTO: A new concept for solar radio spectrometers. Sol. Phys. 226(1). 143–151. doi: 10.1007/s11207-005-5688-9

Carley, E. P., Vilmer, N., Simões, P. J. A. *et al*., 2017. Estimation of a coronal mass ejection magnetic field strength using radio observations of gyrosynchrotron radiation. Astron. Astrophys. 608, A137, 1-14. doi: 10.1051/0004-6361/201731368

Cho, K. S., Gopalswamy, N., Kwon, R. Y. *et al*., 2013. A high-frequency type II solar radio burst associated with the 2011 february 13 coronal mass ejection. Astrophys. J. 765(2), 148, 1-9. https://doi.org/10.1088/0004-637X/765/2/148

Cho, K. S., Lee, J., Moon, Y. J. *et al*., 2007. A study of CME and type II shock kinematics based on coronal density measurement. Astron. Astrophys. 461(3), 1121–1125. https://doi.org/10.1051/0004-6361:20064920

Clette, F., Lefèvre, L., Cagnotti, M. *et al*., 2016. The Revised Brussels–Locarno Sunspot Number (1981 – 2015). Sol. Phys. 291(9), 2733–2761. https://doi.org/10.1007/s11207-016-0875-4

Cliver, E. W., Nitta, N. V., Thompson, B. J. *et al*., 2005. Coronal shocks of november 1997 revisited: the cme-type ii timing problem. Sol. Phys. 225(1), 105-139.

Cunha-Silva, R. D., Fernandes, F. C. R., Selhorst, C. L., 2014. Solar type II radio bursts recorded by the compound astronomical low-frequency low-cost instrument for spectroscopy in transportable observatories in Brazil. Sol. Phys. 289(12), 4607–4620. https://doi.org/10.1007/s11207-014-0586-7

Gopalswamy, N., 2011. Coronal mass ejections and solar radio emissions. Planetary Radio Emissions, 7, 325–342. https://doi.org/10.1553/pre7s325

Gopalswamy, N., Akiyama, S., Makela, P. *et al*., 2016. On the directivity of low-frequency type IV radio bursts. 2016 URSI Asia-Pacific Radio Science Conference, URSI AP-RASC 2016, 1247–1249. https://doi.org/10.1109/URSIAP-RASC.2016.7601385

Gopalswamy, N., Xie, H., Mäkelä, P. *et al.*, 2013. Height of shock formation in the solar corona inferred from observations of type II radio bursts and coronal mass ejections. Adv. Space Res. 51(11), 1981–1989. https://doi.org/10.1016/j.asr.2013.01.006

Gopalswamy, N., Yashiro, S., Akiyama, S. *et al*., 2008. Coronal mass ejections, type II radio bursts, and solar energetic particle events in the SOHO era. Ann. Geophys. 26(10), 3033–3047. https://doi.org/10.5194/angeo-26-3033-2008

Gopalswamy, N., Kaiser, M. L., Lepping, R. P. *et al*., 1998. Origin of coronal and interplanetary shocks: A new look with WIND spacecraft data. J. Geophys. Res, 103(A1), 307-316. https://doi.org/10.1029/97JA02634

Gopalswamy, N., Nitta, N.V., Akiyama, S. *et al*., 2012. Coronal magnetic field measurement from EUV images made by the solar dynamics observatory. Astrophys. J. 744(1), 72, 1-7. https://doi.org/10.1088/0004-637X/744/1/72

Grechnev, V. V., Uralov, A. M., Chertok, I. M. *et al*., 2011. Coronal shock waves, EUV waves, and their relation to CMEs. I. Reconciliation of "EIT waves", Type II radio bursts, and leading edges of CMEs. Sol. Phys. 273, 433–460. https://doi.org/10.1007/s11207-011-9780-z

Hariharan, K., Ramesh, R., Kishore, P. *et al*., 2014. An estimate of the coronal magnetic field near a solar coronal mass ejection from low-frequency radio observations. Astrophys. J. 795(1), 14, 1-9. https://doi.org/10.1088/0004-637X/795/1/14

Huang, W., Aa, E., Shen, H. *et al*., 2018. Statistical study of GNSS L-band solar radio bursts. GPS Sol. 22(4),114, 1-9. https://doi.org/10.1007/s10291-018-0780-4

Kishore, P., Ramesh, R., Hariharan, K. *et al*., 2016. Constraining the solar coronal magnetic field strength using split-band type II radio burst observations. Astrophys. J. 832(1), 59, 1-7. https://doi.org/10.3847/0004-637x/832/1/59

Kumari, A., Morosan, D. E., Kilpua, E. K. J., 2020. On the occurrence of type IV solar radio bursts in the solar cycle 24 and their association with coronal mass ejections. Astrophys. J. 906(2), 79, 1-9. https://doi.org/10.3847/1538-4357/abc878

Kumari, A., Ramesh, R., Kathiravan, C. *et al*., 2017a. New evidence for a coronal mass ejection-driven high frequency type II burst near the sun. Astrophys. J. 843(1), 10, 1-7. https://doi.org/10.3847/1538-4357/aa72e7

Kumari, A., Ramesh, R., Kathiravan, C. *et al*., 2017b. Strength of the solar coronal magnetic field – A comparison of independent estimates using contemporaneous radio and white-light observations. Sol. Phys. 292(11), 1–15. https://doi.org/10.1007/s11207-017-1180-6

Kumari, A., Ramesh, R., Kathiravan, C. *et al*., 2019. Direct estimates of the solar coronal magnetic field using contemporaneous extreme-ultraviolet, radio, and white-light observations. Astrophys. J. 881(1), 24, 1-8. https://doi.org/10.3847/1538-4357/ab2adf

Lata Soni, S., Ebenezer, E., Lal Yadav, M., 2021. Multi-wavelength analysis of CME-driven shock and type II solar radio burst band-splitting. Astrophys. Space Sci. 366(31), 1- 10. https://doi.org/10.1007/s10509-021-03933-7

Lobzin, V., Cairns, I. H., Robinson, P. A., 2011. Solar cycle variations of the occurrence of coronal type III radio bursts and a new solar activity index. Astrophys. J. Lett. 736(1), L20, 1-7. https://doi.org/10.1088/2041-8205/736/1/L20

Mahender, A., Sasikumar Raja, K., Ramesh, R. *et al*., 2020. A statistical study of low-frequency solar radio type III bursts. Sol. Phys. 295(11), 153, 1-10. https://doi.org/10.1007/s11207-020-01722-z

Majumdar, S., Tadepalli, S. P., Maity, S. S. *et al*., 2021. Imaging and spectral observations of a type II radio burst revealing the section of the CME-driven shock that accelerates electrons. Sol. Phys. 296(62), 1-16. https://doi.org/10.1007/s11207-021-01810-8

Mann, G., Classen, T., Aurass, H., 1995. Characteristics of coronal shock waves and solar type II radio bursts. Astron. Astrophys. 295, 775–781.

Mercier, C., Subramanian, P., Chambe, G. *et al*., 2015. The structure of solar radio noise storms. Astron. Astrophys. 576, A136, 1-10. https://doi.org/10.1051/0004-6361/201321064

Morosan, D. E., Kumari, A., Kilpua, E. K. J. *et al*., 2021. Moving solar radio bursts and their association with coronal mass ejections. Astron. Astrophys. 647, 1–5. https://doi.org/10.1051/0004-6361/202140392

Morosan, D. E., Palmerio, E., Räsänen, J. E. *et al*., 2020. Electron acceleration and radio emission following the early interaction of two coronal mass ejections. Astron. Astrophys. 642, 1–13. https://doi.org/10.1051/0004-6361/202038801

Mujiber Rahman, A., Umapathy, S., Shanmugaraju, A. *et al.*, 2012. Solar and interplanetary

parameters of CMEs with and without type II radio bursts. Adv. Space Res. 50(4), 516–525. https://doi.org/10.1016/j.asr.2012.05.003

Nakagawa, Y., Nozawa, S., Shinbori, A., 2019. Relationship between the low-latitude coronal hole area, solar wind velocity, and geomagnetic activity during solar cycles 23 and 24. Earth, Planets and Space, 71(1), 24, 1-15. doi: 10.1186/s40623-019-1005-y

Ndacyayisenga, T., Uwamahoro, J., Raja, K. S. *et al*., 2020. A Statistical Study of Solar Radio Type III Bursts and Space Weather Implication. Adv. Space Res. 67(4), 1425–1435. https://doi.org/10.1016/j.asr.2020.11.022

Nedal, M., Mahrous, A., Youssef, M., 2019. Predicting the arrival time of CME associated with type II radio burst. Astrophys. Space Sci. 364(161), 1-11. https://doi.org/10.1007/s10509-019-3651-8

Newkirk, Jr. G., 1961. The solar corona in active regions and the thermal origin of the slowly varying component of solar radio radiation. Astrophys J. 133, 983-1013.

Pohjolainen, S., Van Driel-Gesztelyi, L, Culhane, J. L. *et al*., 2007. CME Propagation characteristics from radio observations. Sol. Phys. 244, 167–188. https://doi.org/10.1007/s11207-007-9006-6

Ramesh, R., Kathiravan, C., Kartha, S. S. *et al*., 2010. Radioheliograph observations of metric type ii bursts and the kinematics of coronal mass ejections. Astrophys. J. 712(1), 188–193. https://doi.org/10.1088/0004-637X/712/1/188

Ramesh, R., Raja, K. S., Kathiravan, C. *et al.*, 2013. Low-frequency radio observations of picoflare category energy releases in the solar atmosphere. Astrophys. J. 762(2), 89, 1-6. https://doi.org/10.1088/0004-637X/762/2/89

Ratcliffe, H., Kontar, E. P., Reid, H. A. S., 2014. Large-scale simulations of solar type III radio bursts: Flux density, drift rate, duration, and bandwidth. Astron. Astrophys. 572, A111, 1–11. https://doi.org/10.1051/0004-6361/201423731

Sasikumar Raja, K., Subramanian, P., Ananthakrishnan, S. et al., 2018. CALLISTO Spectrometer at IISER-Pune, *ArXiv e-prints*, 1-15. https://ui.adsabs.harvard.edu/link_gateway/2018arXiv180103547S/arxiv:1801.03547

Schmidt, J. M., Cairns, I. H., Lobzin, V. V., 2014. The solar-type II radio bursts of 7 March 2012: Detailedsimulation analyses. J. Geophys. Res: Space Phys. 119(8), 6042–6061. https://doi.org/10.1002/2014JA019950

Shanmugaraju, A., Bendict Lawrance, M., Moon, Y. J. *et al*., 2017. Heights of coronal mass ejections and shocks inferred from metric and DH type II radio bursts. Sol Phys, 292(9). 1-15. https://doi.org/10.1007/s11207-017-1155-7

Singh, D., Sasikumar Raja, K., Subramanian, P. *et al*., 2019. Automated detection of solar radio bursts using a statistical method. Sol Phys. 294(8). 1-14. https://doi.org/10.1007/s11207-019-1500-0

Smerd, S.F., Sheridan, K.V., Stewart, R.T., 1975. Split-band structure in type II radio bursts from the sun. Astrophys lett.16, 23-28.

Temmer, M., Veronig, A. M., Kontar, E. P. *et al*., 2010. Combined STEREO/RHESSI study of coronal mass ejection acceleration and particle acceleration in solar flares. Astrophys. J. 712(2), 1410–1420. https://doi.org/10.1088/0004-637X/712/2/1410

Tidman, D. A. (1965). Radio emission from shock waves and type II solar outbursts. Planet. Space Sci. 3(8), 781–788. https://doi.org/10.1016/0032-0633(65)90115-7

Umuhire, A. C., Gopalswamy, N., Uwamahoro, J. *et al*., 2021a. Properties of high-frequency type II radio bursts and their relation to the associated coronal mass ejections. Sol Phys. 296(27),

1-18. https://doi.org/10.1007/s11207-020-01743-8

Umuhire, A. C., Uwamahoro, J., Sasikumar Raja, K. *et al.*, 2021b. Trends and characteristics of high-frequency type II bursts detected by CALLISTO spectrometers. Adv. Space Res. 68(8), 3464–3477. https://doi.org/10.1016/j.asr.2021.06.029

Vasanth, V., Umapathy, S., 2013. A statistical study on CMEs associated with DH-type-II radio bursts based on their source location (limb and disk events). Sol. Phys. 282(1), 239–247. https://doi.org/10.1007/s11207-012-0126-2

Vasanth, V., Umapathy, S., Vršnak, B. *et al.*, 2011. Characteristics of type-II radio bursts associated with flares and CMEs. Sol. Phys. 273(1), 143–162. https://doi.org/10.1007/s11207-011-9854-y

Vasanth, V., Umapathy, S., Vršnak, B. *et al.*, 2014. Investigation of the coronal magnetic field using a type II solar radio burst. Sol. Phys. 289(1), 251–261. https://doi.org/10.1007/s11207-013-0318-4

Vršnak, B., Magdalenić, J., Aurass, H., 2001a. Comparative analysis of type ii bursts and of thermal and non-thermal flare signatures. Sol. Phys. 202(2), 319-335.

Vr˘snak, B., Aurass, H., Magdalenić, J. et al., 2001b. Band-splitting of coronal and interplanetary type II bursts I: Basic properties. Astron. Astrophys. 377(3), 321–329. https://doi.org/10.1051/0004-6361:20011067

Vršnak, B., Magdalenić, J., Aurass, H. *et al.*, 2002. Band-splitting of coronal and interplanetary type II bursts II: Coronal magnetic field and Alfvén velocity. Astron. Astrophys. 396(2), 673–682. https://doi.org/10.1051/0004-6361:20021413

Zimovets, I. V., Sadykov, V. M., 2015. Spatially resolved observations of a coronal type II radio burst with multiple lanes. Adv. Space Res. 56(12), 2811–2832. https://doi.org/10.1016/j.asr.2015.01.041

Zucca, P., Carley, E. P., Mccauley, J. *et al.*, 2012. Observations of low frequency solar radio bursts from the Rosse solar-terrestrial observatory. Sol. Phys. 280(2), 591-602. https://doi.org/10.1007/s11207-012-9992-x